\documentclass[aps,twocolumn,prl,amsmath,amssymb,showpacs,superscriptaddress]{revtex4}
\usepackage{graphicx,multirow,array}

\begin{document}
\title{The spin glass transition of the three dimensional Heisenberg
spin glass.}

\author{I. Campos} \affiliation{Instituto de Biocomputaci\'on y F\'{\i}sica de
  Sistemas Complejos (BIFI), Corona de Arag\'on 42, Zaragoza 50009, Spain.}

\author{M. Cotallo-Aban} \affiliation{Departamento de F\'{\i}sica Te\'orica,
  Universidad de Zaragoza, Zaragoza 50009, Spain.}  \affiliation{Instituto de
  Biocomputaci\'on y F\'{\i}sica de Sistemas Complejos (BIFI), Corona de
  Arag\'on 42, Zaragoza 50009, Spain.}

\author{V. Martin-Mayor}
\affiliation{Departamento de F\'{\i}sica Te\'orica I, Facultad de
Ciencias F\'{\i}sicas, Universidad Complutense, 28040 Madrid, Spain.}
\affiliation{Instituto de Biocomputaci\'on y F\'{\i}sica de Sistemas
Complejos (BIFI), Corona de Arag\'on 42, Zaragoza 50009, Spain.}

\author{S. Perez-Gaviro} \affiliation{Departamento de F\'{\i}sica
Te\'orica, Universidad de Zaragoza, Zaragoza 50009, Spain.}
\affiliation{Instituto de Biocomputaci\'on y F\'{\i}sica de Sistemas
Complejos (BIFI), Corona de Arag\'on 42, Zaragoza 50009, Spain.}

\author{A. Tarancon} \affiliation{Departamento de F\'{\i}sica Te\'orica,
  Universidad de Zaragoza, Zaragoza 50009, Spain.} \affiliation{Instituto de
  Biocomputaci\'on y F\'{\i}sica de Sistemas Complejos (BIFI), Corona de
  Arag\'on 42, Zaragoza 50009, Spain.}

\date{\today}
\begin{abstract}
It is shown, by means of Monte Carlo simulation and Finite Size
Scaling analysis, that the Heisenberg spin glass undergoes a
finite-temperature phase transition in three dimensions. There is a
single critical temperature, at which both a spin glass and a chiral
glass orderings develop. The Monte Carlo algorithm, adapted from
lattice gauge theory simulations, makes possible to thermalize lattices
of size $L=32$, larger than in any previous spin glass simulation in
three dimensions. High accuracy is reached thanks to the use of the
{\em Marenostrum} supercomputer. The large range of system sizes
studied allow us to consider scaling corrections.
\end{abstract}
\pacs{
75.50.Lk  
75.40.Mg. 
64.60.Fr, 
05.50.+q, 
}

\maketitle

Recently, decisive evidence for a spin glass transition~\cite{SGINTRO}
in three dimensional Ising spin glasses was found by
us~\cite{BALLESTEROS} and by Palassini and
Caracciolo~\cite{PALASS-CARACC}. These works demonstrated the
applicability to spin-glasses of our
approach~\cite{QUOTIENTS,VICTORAMIT} to Finite Size Scaling (FSS) at
the critical temperature, as well as that of Caracciolo and
coworkers~\cite{FSS-PISA} for the paramagnetic state (see
also~\cite{JORG,KATZYOUNG}).

However, the situation is still confusing for {\em Heisenberg} spin
glasses~\cite{Tc0,MATSUBARA,IMAGAWA,VILLAIN,QUIRALES,QUIRALESNEW,YOUNG,OTROS,FELIX},
which is the more experimentally interesting case (see
e.g.~\cite{MEM-REJ-BERT,HEISS-H-EXP}). Indeed, numerical studies are
the only theoretical tool to achieve progress in three
dimensions. Early simulations~\cite{Tc0} could not reach low enough
temperatures due to the dramatic dynamical arrest of the algorithms
available at the time, and concluded that the critical temperature,
$T_\mathrm{c}$, was strictly zero.  Yet, recent studies at lower
temperatures~\cite{OTROS,FELIX} found indirect indications of a spin
glass transition for Heisenberg spin glasses.  Matters are further
complicated by the Villain et al.~\cite{VILLAIN} suggestion of a low
temperature chiral glass phase, with an Ising like ordering due to the
handedness of the non-collinear spin structures (see definitions
below). The simulations of Kawamura and
coworkers~\cite{QUIRALES,QUIRALESNEW} gave ample support to this
spin-chirality decoupling scenario (i.e. $T_\mathrm{c}=0$ for the spin
glass, but $T_\mathrm{c}>0$ for the chiral glass ordering).

In order to clarify the situation for Heisenberg and XY spin glasses
in $D=3$, Young and Lee~\cite{YOUNG} have recently tried our FSS
methods at $T_\mathrm{c}$~\cite{BALLESTEROS,QUOTIENTS}.  Although
parallel tempering only allowed them~\cite{YOUNG} to thermalize
systems of size up to $L=12$, very clear results were reached for the
XY spin glasses. The finite-lattice correlation length~\cite{COOPER},
$\xi_L$, was analyzed for several system sizes $L$.  As
expected~\cite{QUOTIENTS,BALLESTEROS,VICTORAMIT}, the dimensionless
ratio $\xi_L/L$ crosses neatly at the {\em same} $T_\mathrm{c}$, for
the chiral glass and the spin glass ordering, for XY spin glasses. In
the more important case of Heisenberg spin glasses, their results,
although inconclusive, were interpreted also as lack of spin-chirality
decoupling. This conclusion has been criticized by Kawamura and
Hukushima~\cite{QUIRALESNEW}, that studied somehow larger systems on
{\em very} few samples.

Here we show that a finite-temperature spin glass transition occurs
for the Heisenberg spin glass in $D=3$. The critical temperature for
the spin glass transition coincides with that of the chiral glass. Our
results rely on Monte Carlo simulation and FSS analysis at
$T_\mathrm{c}$~\cite{QUOTIENTS,BALLESTEROS,VICTORAMIT}. We adapt a
lattice gauge theory algorithm~\cite{OVERRELAX} to our problem. Our
algorithm thermalizes $L=32$ systems, well beyond any previous spin
glass simulation in $D=3$. The use of {\em Marenostrum}, one of the
World largest computing facilities, during (the equivalent of)
$2.6\times 10^4$ Pentium IV computing days allowed us to simulate 4000
samples. The phase transition seems to be of Kosterlitz-Thouless
type~\cite{KOST-THOU}, although we may not exclude a lower critical
dimension barely smaller than three.


We consider the Edwards-Anderson model.  The Heisenberg spins
$\vec{S_i}=(S_i^x, S_i^y, S_i^z)$, $\vec{S_i}\cdot\vec{S_i}=1\,$, live
on the nodes a cubic lattice of size $L$, with periodic boundary
conditions.  Spins interact via the Hamiltonian ($\langle ij\rangle$
indicates sum over all pairs of lattice nearest neighbors):
\begin{equation}
  H=- \sum_{\langle i,j\rangle } J_{ij} \vec{S_i} \cdot
  \vec{S_j}\,.\label{HAMILTONIAN}
\end{equation}
The $J_{ij}$ are Gaussian distributed quenched random
couplings~\cite{SGINTRO}. Their mean value is zero, while their
variance sets the energy unit. For any quantity, $O$, we first
calculate the thermal average for the given couplings, $\langle
O\rangle_J$. The average over the $J_{ij}$, $\overline{\langle
O\rangle_J}$, is only taken afterward.

In order to detect non-planar spin structures, one forms the chiral
(pseudovector) density field~\cite{QUIRALES}:
\begin{equation}
  \zeta_{i\mu}= \vec{S}_{i+\hat{e}_{\mu}} \cdot ( \vec{S}_{i} \times
  \vec{S}_{i-\hat{e}_{\mu}})\quad (\mu=x,y,z)\,,\label{LOCALQUIRAL}
\end{equation}
where $\hat{e}_{\mu}$ is the unit lattice vector along the $\mu$ axis.

We introduce real replicas~\cite{SGINTRO}: pairs of spin
configurations, $\vec S_i^{(1)}$ and $\vec S_i^{(2)}$ independently
evolving with the same set of couplings, $\{J_{ij}\}$, and at the same
temperature, $T$. We have as well a replicated field for the chiral
densities, (\ref{LOCALQUIRAL}).  From the real replicas we form the
tensor field $q_{\mu\nu}(\vec r_i)$ and the chiral vector field
$\kappa_{i,\mu}$: ($\mu,\nu=x,y,z$)
\begin{equation}
q_{\mu, \nu}(\vec r_i)= S_{i,\mu}^{(1)} S_{i,\nu}^{(2)}\,,\quad\kappa_{i,\mu}=\zeta_{i,\mu}^{(1)}\zeta_{i,\mu}^{(2)}\,.
\end{equation}
Their Fourier transforms,
\begin{equation}
\hat q_{\mu, \nu}(\vec k)= \sum_{i=1}^N \frac{q_{\mu,\nu}(\vec r_i)
\mathrm{e}^{\mathrm{i} \vec{k} \cdot \vec{r_i}}}{L^3} \,,\ 
\hat\kappa_\mu(\vec k)= \sum_{i=1}^N\frac{\kappa_{i,\mu}
\mathrm{e}^{\mathrm{i} \vec{k} \cdot \vec{r_i}}}{L^3}\,,
\end{equation}
yield the spin glass (SG) wave vector dependent susceptibility and the
chiral glass (CG) one:
\begin{equation}
\frac{\chi_\mathrm{SG}(\vec{k})}{L^3}=\sum_{\mu,\nu} \overline{\langle |\hat
      q_{\mu, \nu}(\vec k)|^2 \rangle_J}\,,\  
\frac{\chi_\mathrm{CG}^\mu(\vec{k})}{L^3}=\overline{\langle |\hat \kappa_{\mu}(\vec k)|^2
\rangle_J}.\label{SG_CG_SUSCEPT}
\end{equation}

Both for the SG and the CG case, we have~\cite{COOPER}:
\begin{equation}
\xi_L= \frac{1}{2 \sin(k_\mathrm{min}/2) } \left(
\frac{\chi(\vec{0})}{\chi(\vec{k}_\mathrm{min})} - 1 \right)^{1/2},
\end{equation}
where $\vec k_\mathrm{min}=(2\pi/L,0,0)$ or permutations.  One has
$\xi^{\bot}_\mathrm{CG}$ or $\xi^{\Vert}_\mathrm{CG}$ for $\hat
e_\mu\cdot\vec{k}_\mathrm{min}=0$ or $2\pi/L$,
respectively~\cite{YOUNG}.  Rotational invariance at $T_\mathrm{c}$
implies $\xi^{\bot}_\mathrm{CG}=\xi^{\Vert}_\mathrm{CG}$ for large
$L$.

\begin{figure}[t]
\begin{center}
\includegraphics[width=\columnwidth ]{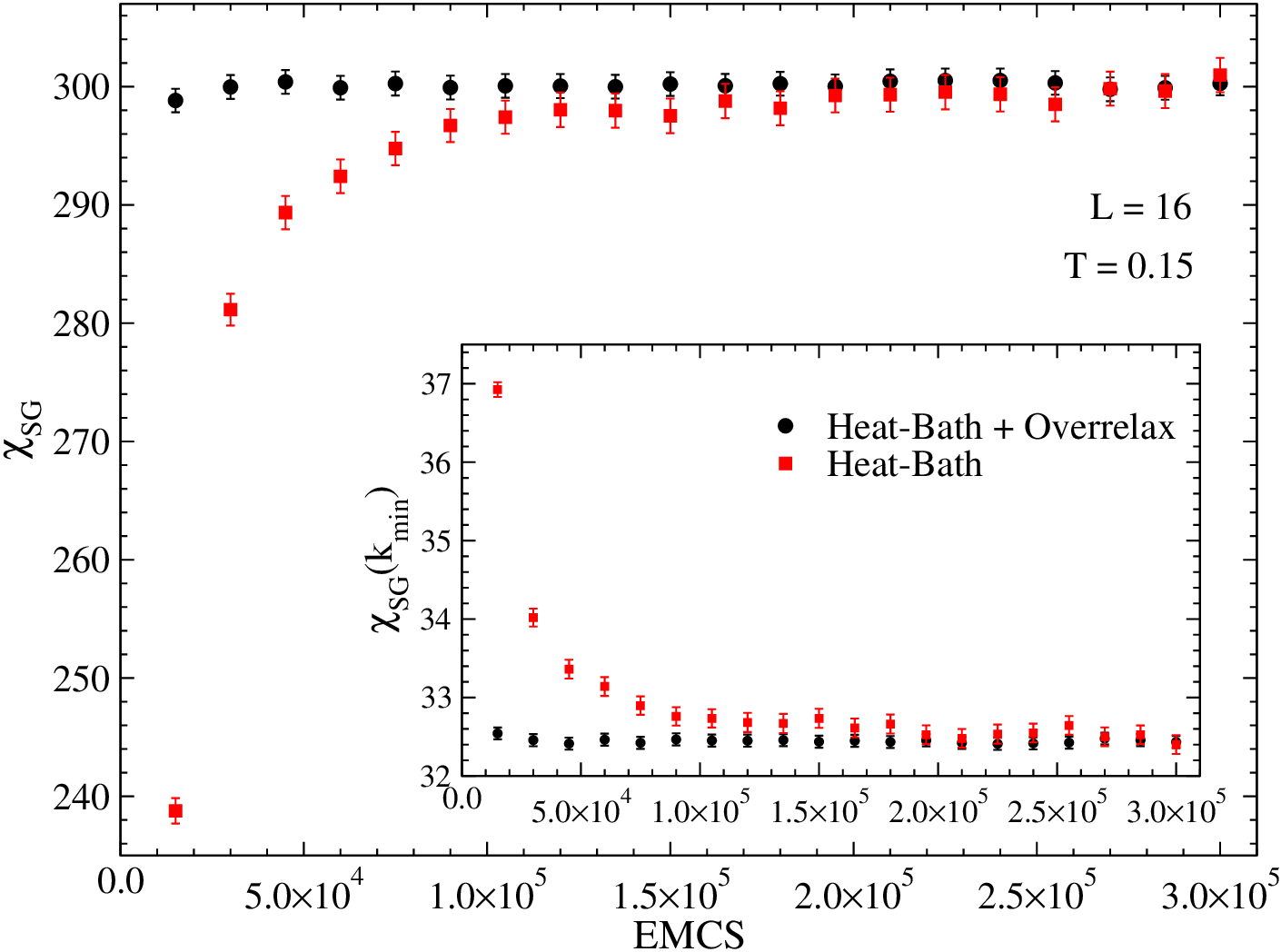}
\caption{Comparison of heat bath vs. heat bath plus overrelaxation
  evolution of $\chi_\mathrm{SG}(\vec 0)$ ( $\chi_\mathrm{SG}(\vec
  k_{min})$ in the inset) from a hot start (EMCS: either 1 heat bath
  step followed by 20 overrelaxation sweeps or 4 heat bath sweeps).
  Points in the plot are the average over 2500 samples of 15000
  successive EMCS.}
\label{HB_vs_HB-OV}
\end{center}
\end{figure}

Model (\ref{HAMILTONIAN}) was simulated with a mixture of heat bath
and (microcanonical) overrelaxation taken from lattice
QCD~\cite{OVERRELAX} but also effective for frustrated spin
models~\cite{OVERRELAXSPIN}.  We straightforwardly modified the
implementation for $J_{ij}=1$ in~\cite{VICTORAMIT}. The Elementary
Monte Carlo Step (EMCS) consists on a sequential heat bath sweep,
followed by (the integer part of) $5L/4$ {\em sequential}
overrelaxation sweeps, to let the microcanonical wave run over the
system.  Overrelaxation fastly evolves chirality (in $D=1$ it inverts
the local chirality).  One heat bath update is roughly as CPU time
consuming as 7 overrelaxations. The merits of the mixed algorithm can
be assessed from Fig.~\ref{HB_vs_HB-OV} (standard Parallel Tempering
reached only $L=12$ for the same model~\cite{YOUNG}). Overrelaxation
may be combined with Parallel Tempering, but this is unnecessary at
$T_\mathrm{c}$. Lattices $L=4,6,8,12,16,24$ and $32$ were simulated
close to $T_\mathrm{c}\approx 0.16(2)$~\cite{YOUNG,QUIRALES} (see
table~\ref{SIMUDETAILS}). We extrapolate to nearby temperatures using
bias-corrected~\cite{BIAS} data reweighting~\cite{SPECTRALDENSITY}.

\begin{table}[t]
\begin{center}
\begin{tabular}{|c|r|r|r|r|r|r|r|}
\hline 
L & $4$ & $6$ & $8$ & $12$ & $16$ & $24$ & $32$\\ 
\hline
T &0.15 &0.15 & 0.15&0.150& 0.150& 0.146 & 0.145\\
    & 0.16& 0.16& 0.155&0.155& 0.156& 0.150 & 0.147\\
    & & & 0.160&0.160& 0.160& 0.155 & 0.150\\
\hline 
$N_\mathrm{samples} $& 2& 4& 2&3 & 2.5& 2& $1$\\ 
$(\times 10^{3})$ & 2& 6& 2&2.4& 2.5& 1.3& $1$\\
                  & & & 2& 3& 2& 2& $2$\\ 
\hline EMCS$\times 10^{5}$& $1$ &$1$ &$1$ &$2$&$3$ &$16$ & $40$\\ 
\hline
\end{tabular}
\end{center}
\caption{{\bf Details of simulation.} For each lattice size, we give
the studied temperatures, the number of simulated samples (in the same
order), and the number of elementary Monte Carlo steps per sample. In
each sample we took $10^5$ measurements, excepting for $L=32$ (20000
measurements). }
\label{SIMUDETAILS}
\end{table}


Thermalization is a major issue in SG simulations.  Our thermalization
tests included the by now standard $\log_2$ data binning (i.e. average
over all samples the second half of the generated data, and compare
this average with that of the second fourth of the Monte Carlo
history, the average over the second eighth, and so on), finding
compatibility for the last three bins.  Another strong thermalization
test is the consistency of the reweighting
extrapolation~\cite{SPECTRALDENSITY}. As Fig.~\ref{XISG_over_L} shows,
the reweighting extrapolation is satisfactory for our data. It is a
nice check, because simulations at different $T$ are completely
independent.

We use the quotients FSS method~\cite{QUOTIENTS}. The used scaling
variable is $\xi_L/L$~\cite{FSS-PISA,QUOTIENTS}, rather than the
unknown $L/\xi_\infty$ or $(T-T_\mathrm{c})L^{1/\nu}$.  For an
observable $O$, diverging in the large $L$ limit as $\overline{\langle
O\rangle}_{J,L=\infty}\propto |T-T_\mathrm{c}|^{x_O}$, we compare
$\overline{\langle O\rangle_J}$ in lattices $L_1$ and $L_2$, at the
crossing temperature $T^{L_1,L_2}$ such that
$\xi_{L_1}(T^{L_1,L_2})/L_1=\xi_{L_2}(T^{L_1,L_2})/L_2\,$:
\begin{equation}
\frac{\overline{\langle O_{L_2}(T^{L_1,L_2})
    \rangle_J}}{\overline{\langle O_{L_1}(T^{L_1,L_2}) \rangle_J}}=
    \left( \frac{L_2}{L_1} \right)^{\frac{x_O}{\nu}} +\ldots\,,
\label{QUOTIENTS-FORMULA}
\end{equation}
where $\nu$ is the correlation length critical exponent (the dots
stand for scaling corrections~\cite{VICTORAMIT}). The advantages of
Eq.(\ref{QUOTIENTS-FORMULA}) are
many~\cite{QUOTIENTS,BALLESTEROS,BIAS,JORG}. It is easy to use.
Arbitrary choices on the temperature range are avoided. The
statistical error estimation is crystal clear. One directly observes
scaling corrections by increasing $L_1$ and $L_2$.

Should $T^{L_1,L_2}$ be obtained from $\xi_{L,\mathrm{CG}}$ or from
$\xi_{L,\mathrm{SG}}$?  There is a consensus on the divergence of
$\xi_{\mathrm{CG}}$ at $T_\mathrm{c}$, while that of $\xi_\mathrm{SG}$
is controversial~\cite{YOUNG,QUIRALESNEW}.  Moreover,
$\xi_{L,\mathrm{CG}}$ suffers smaller scaling corrections
(Fig.~\ref{FIG4}). So, the results in table~\ref{table_exp_1_over_nu}
were obtained with $\xi_{L,\mathrm{CG}}$.  As for
$\xi_{L,\mathrm{SG}}$, we treat it here as an observable scaling as
$L^{\phi}$.  Some groups~\cite{YOUNG,OTROS,FELIX} expect $\phi=1$,
while others~\cite{QUIRALES,QUIRALESNEW} believe that $\phi=0$.

Several features are salient in table~\ref{table_exp_1_over_nu}. (i)
Once we neglect $L=4$, the $L$ evolution of our estimates is monotonic
with increasing $L_1$ and $L_2$ (we agree with
Refs.~\cite{YOUNG,QUIRALESNEW} within their $L$ window). (ii) The
$1/\nu$ from $\partial_T \xi_\mathrm{SG}$ and $\partial_T
\xi_\mathrm{CG}$ is compatible for $L\geq 12$ . (iii) When $L_1$ and
$L_2$ increases, $1/\nu$ systematically decreases (in a
Kosterlitz-Thouless scenario it tends to zero). (iv) Exponent
$\gamma_\mathrm{SG}/\nu$ stabilizes for $L\geq 12$ in a typical $D=3$
value~\cite{QUOTIENTS}, hence:
\begin{equation}
\chi_\mathrm{SG} \propto \xi_\mathrm{CG}^{1.93\pm 0.02}\,.\label{EXPONENTE}
\end{equation}
The scenario of Refs.~\cite{QUIRALES,QUIRALESNEW}, where
$\xi_\mathrm{CG}$ diverges at $T_\mathrm{c}$, while $\xi_\mathrm{SG}$
(and then $\chi_\mathrm{SG}$) does not, becomes untenable.

\begin{table}[b]
\begin{center}
\begin{tabular}{| c  c | c | c | c | c | c | c |}
\hline
$L_1$ & $L_2$ & $\phi$ & $1/\nu_\mathrm{SG}$ & $1/\nu_\mathrm{CG}$ 
&  $\gamma_\mathrm{SG}/\nu_\mathrm{SG}$ &  $\gamma_\mathrm{CG}/\nu_\mathrm{CG}$\\
\hline
\hline
4 & 8& 1.047(5) & $ 1.278(10) $ & $ 1.077(20) $ &  $ 2.117(7) $ & $ 1.061
(10) $\\
\hline
6 & 12& 1.057(3) & $ 1.223(10) $ & $ 0.761(12) $ &  $ 2.056(5) $ & $ 0.628 
(11) $\\
\hline
8 & 16& 1.013(4) & $ 0.989(10) $ & $ 0.722(12) $ &  $ 2.008(8) $ & $ 0.738
(18) $\\
\hline
12 & 24& 0.939(4) & $ 0.71(17) $ & $ 0.66(11) $ &  $ 1.954(9) $ & $ 0.925
(22) $\\
\hline
16 & 24& 0.917(5) & $ 0.68(11) $ & $ 0.67(12) $ &  $ 1.943(15) $ & $ 1.02
(4) $\\
\hline
16 & 32& 0.908(5) & $ 0.72(20) $ & $ 0.73(11) $ &   $ 1.945(13) $ & $ 1.130
(29) $\\
\hline
24 & 32$^A$& 0.88(2) & $ 0.60(10) $ & $ 0.64(12) $ &  $ 1.93(2) $ & $ 1.26(7) 
$\\
\hline
24 & 32$^B$ & 0.91(2) & $0.56(12)$ & $0.51(13)$ &  $1.96(4)$ & $1.36
(11)$\\
\hline
\end{tabular}
\end{center}
\caption{Critical exponents from Eq.(\ref{QUOTIENTS-FORMULA})
($T^{L_1,L2}$ from $\xi_{L,\mathrm{CG}}^\Vert$). The $O$ were:
$\xi_\mathrm{SG}$ ($\phi$: $\xi_\mathrm{SG}\sim L^{\phi}$),
$\partial_T \xi_L/L$ ($1/\nu$) and $\chi$ ($\gamma/\nu$). The
$L_1=24$, $L_2=32$ crossing is unclear (Fig.~\ref{FIG4}), so we quote
results at $T^{L_1,L_2}=0.146$ ($A$) and $0.145$
($B$).}\label{table_exp_1_over_nu}
\end{table}

\begin{figure}[b]
\begin{center}
\includegraphics[width=\columnwidth ]{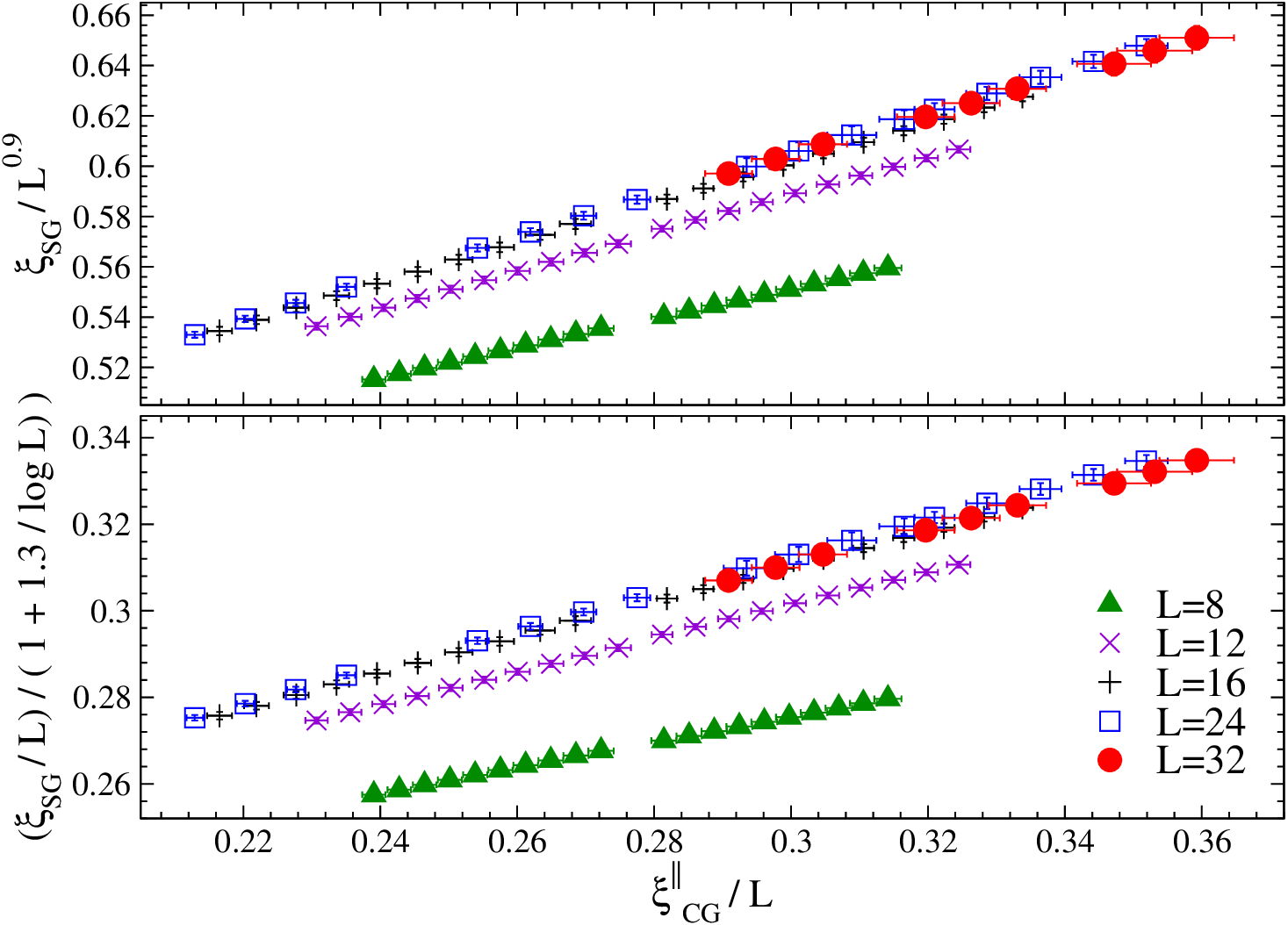}
\caption{(color online) {\bf Top}: For our largest systems
$\xi_\mathrm{SG}$ scales as $L^{0.9}$, using $\xi_\mathrm{CG}/L$ as
scaling variable. {\bf Bottom}: The results in the upper part may be
interpreted as the logarithmic corrections to scaling in
Eq. (\ref{KT-LOG}), particularly for $L>12$.}
\label{CORREC_LOG}
\end{center}
\end{figure}

\begin{figure}
\begin{center}
\includegraphics[width=\columnwidth]{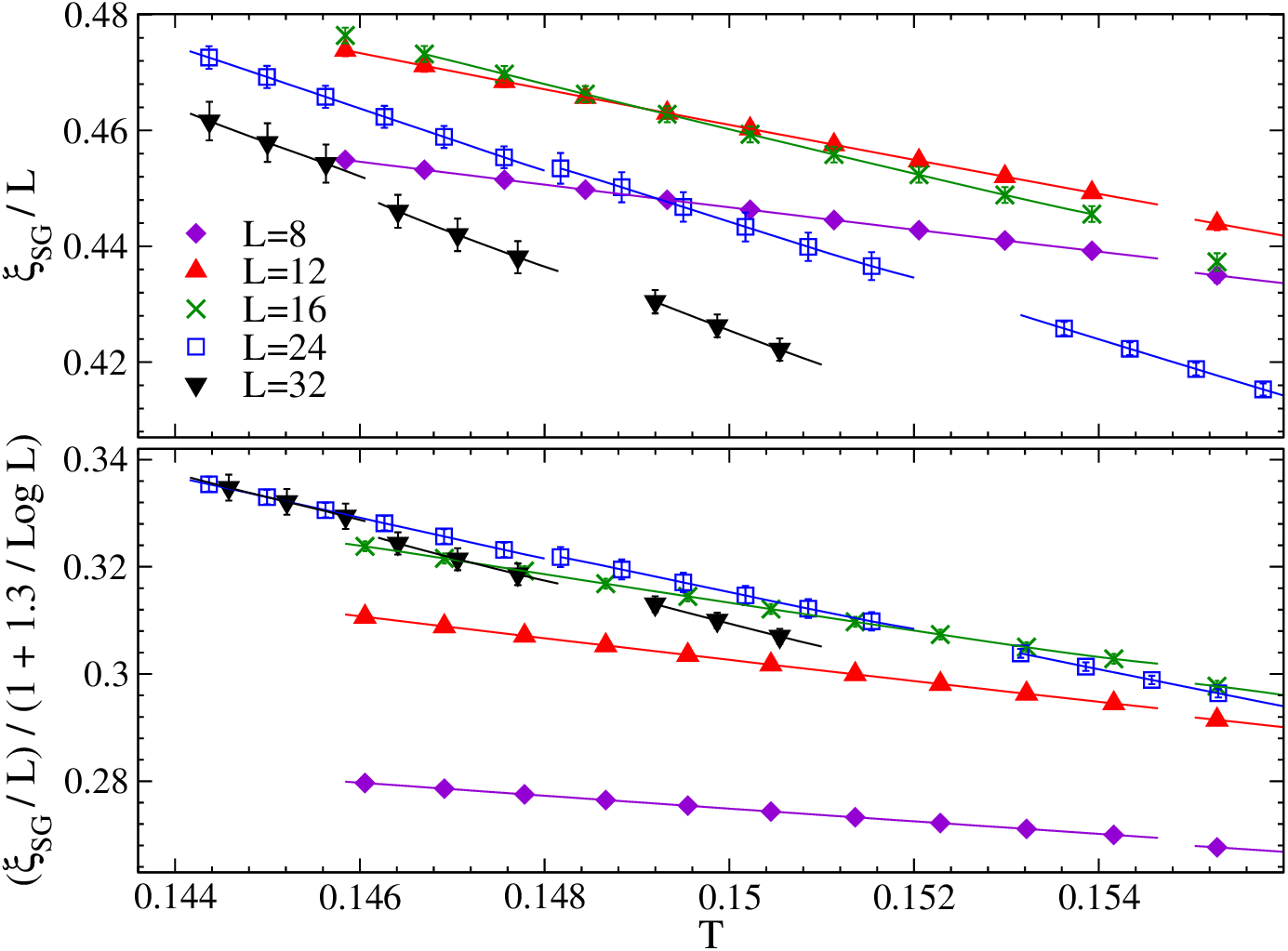}
\caption{(color online) {\bf Top:} $\xi_\mathrm{SG}$ in units of $L$,
versus $T$, for several lattice sizes.  Lines join data obtained from
a single simulation~\cite{SPECTRALDENSITY} (all simulations
independent). For the sake of clarity we only represent data for
$L\geq8$. {\bf Bottom:} as top, correcting the suspected logarithmic
corrections, Eq.(\ref{KT-LOG}).}
\label{XISG_over_L}
\end{center}
\end{figure}

\begin{figure}
\begin{center}
\includegraphics[width=\columnwidth ]{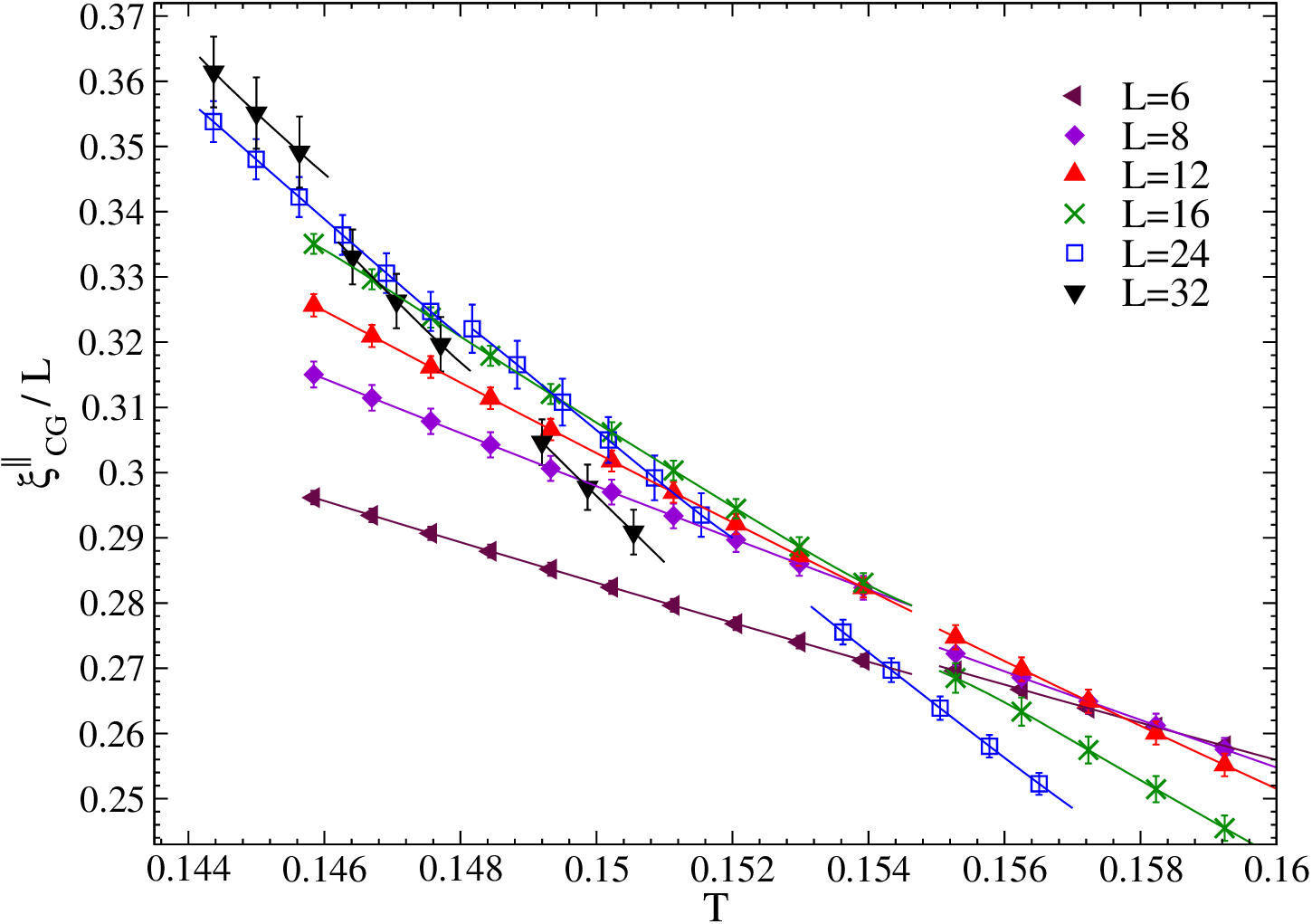}
\caption{(color online) Parallel chiral correlation length in units of
$L$, as a function of $T$, for several lattice sizes.}
\label{FIG4}
\end{center}
\end{figure}

At criticality (table~\ref{table_exp_1_over_nu}), $\xi_\mathrm{SG}$
scales as $L^{0.9}$ rather than $L^1$, immediately suggesting the
presence of logarithmic corrections to scaling.  Without analytical
guidance, the detailed numerical study of logarithmic scaling
corrections~\cite{BIAS} is out of reach. Thus, we borrow the leading
scaling corrections from the $D=2$ XY model~\cite{HASENB} ($f_\xi$ is
a smooth scaling function, while $A$ is a scaling amplitude):
\begin{equation}
\frac{\xi_{L,\mathrm{SG}}}{L}=f_\xi(\xi_{L,\mathrm{CG}}/L)\left(1 +
\frac{A}{\log L}+\ldots\right)\,.
\label{KT-LOG}
\end{equation}
Indeed, see Fig.~\ref{CORREC_LOG}, Eq.(\ref{KT-LOG}) accounts for the
$L^{0.9}$ scaling ($A=1.3$, roughly 4 times that of the 2D-XY
model~\cite{HASENB}). One also expects~\cite{HASENB} multiplicative
logarithms in Eq.(\ref{EXPONENTE}).  Note however that
Eq.(\ref{KT-LOG}) works in a limited range of $L$ for any system
barely above its lower critical dimension since it is the $\omega\to
0$ limit of the standard leading correction to scaling,
$L^{-\omega}$~\cite{VICTORAMIT} (e.g. the $2+\epsilon$
expansion~\cite{2PLUSEPS} yields $\omega\approx\epsilon$ for the
Heisenberg ferromagnet, whose lower critical dimension is two).

We now study $\xi_\mathrm{SG}/L$ as a function of temperature
(Fig. ~\ref{XISG_over_L}).  If we do not correct for scaling
corrections (Fig.~\ref{XISG_over_L}, top), indeed all crossings are in
the rather large range
$T_\mathrm{c}=0.16(2)$~\cite{YOUNG,QUIRALES}. However, we observe a
net shift of the crossings to lower $T$ when the system sizes grow,
(for $L=24$ and $L=32$ a crossing is not found on the simulated $T$
range). Yet, Fig.~\ref{XISG_over_L}--bottom, dividing out the
logarithm scaling corrections in Eq.(\ref{KT-LOG}), the curves for the
largest systems merge around $T=0.146$, as expected for a
Kosterlitz-Thouless transition (see e.g. Ref.~\cite{BALLESTEROS}).

For the sake of completeness, let us consider our data for
$\xi_{L,\mathrm{SG}}$ under the spin-chirality decoupling assumption.
In this scenario, $\xi_{\infty,\mathrm{SG}}$ diverges at $T=0$, as
$\xi_{\infty,\mathrm{SG}}=A/T^\mathrm{\nu^*}$ ($\nu^*=2.2$, according
to~\cite{QUIRALESNEW}). FSS~\cite{VICTORAMIT} predicts that
$\xi_{L,\mathrm{SG}}/L$ is a smooth function of
$L/\xi_{\infty,\mathrm{SG}}$. Take now from
Fig.~\ref{XISG_over_L}--top the temperature where
$\xi_{L,\mathrm{SG}}/L$ reaches (say) 0.44 for $L=24$ and $L=32$,
namely $T_{24}\approx0.1508$ and $T_{32}\approx0.1474$. In other
words, $\xi_{\infty,\mathrm{SG}}$ increases a factor $32/24\approx
1.33$ while the temperature merely decreases by a $2\%$. Matching this
with an algebraic divergence at $T=0$, the nonsensical result
$\nu^*\approx 12$ is obtained.

Consider now $\xi_\mathrm{CG}^\Vert/L$, shown in
Fig.~\ref{FIG4}. Although less than for $\xi_\mathrm{SG}$, the
crossings shift to lower $T$ for larger $L$.  For $L=8,12$ and $16$ we
resolve a crossing at $T\approx 0.155$. For $L\geq 16$ the curves
merge at $T\approx 0.147$, suggesting again a Kosterlitz-Thouless
scaling, but with smaller scaling corrections. The $L=32$ data at low
$T$ are compatible with, but above, the $L=24$ data. We suspect of a
statistical fluctuation (the three $L=32$ simulations are
independent), but a crossing may not be discarded.

In summary, we have shown that the Heisenberg spin glass undergoes a
spin glass transition in $D=3$, by means of Monte Carlo simulation and
FSS analysis~\cite{QUOTIENTS,BALLESTEROS,VICTORAMIT}. We have adapted
a lattice gauge theory algorithm~\cite{OVERRELAX}, that thermalizes
$L=32$ systems, well beyond any previous spin glass simulation in
$D=3$.  Furthermore, a large number of samples were studied in {\em
Marenostrum}.  For studies in the low temperature phase, we suggest to
combine our algorithm with Parallel Tempering. The large range of
system sizes studied allows us to conclude that, at criticality, the
spin glass susceptibility scales with the {\em chiral}
correlation length, Eq.(\ref{EXPONENTE}). Therefore, a single phase
transition is present in this problem: spin-chirality decoupling is
ruled out. We observe logarithmic corrections to scaling, explaining
why previous
investigations~\cite{Tc0,MATSUBARA,IMAGAWA,VILLAIN,QUIRALES,QUIRALESNEW,YOUNG,OTROS,FELIX}
were inconclusive. Our results are compatible with Kosterlitz-Thouless
scaling, but they are typical as well of a system with a lower
critical dimension barely smaller than three.

We thank L. A. Fen\'andez, J.J. Ruiz-Lorenzo and J.L. Alonso for
discussions, and BIFI (U. Z.)  and {\em Marenostrum} (Barcelona
Supercomputing Center) for computing time.  We were partly supported
by BSCH---UCM, by DGA--Spain (M.C.-A and S.P.-G.),  and by MEC (Spain)
through contracts BFM2003-08532, FIS2004-05073.


\end{document}